\documentclass[aps,prl,twocolumn,superscriptaddress,groupedaddress,nofootinbib]{revtex4}  
\usepackage{graphicx}  
\usepackage{dcolumn}   
\usepackage{bm}        
\usepackage{amssymb}   
\graphicspath{{figures/}} 
\usepackage{epstopdf}	
\usepackage{CJK}
\usepackage{amsmath} 

\usepackage{textcomp}

\begin{document}
	
	\begin{CJK*}{}{} 
		\title{High-Precision Ramsey-Comb Spectroscopy Based on High-Harmonic Generation}
		\author{L.S.~Dreissen}
		\author{C.~Roth}
		\author{E.L.~Gr\"{u}ndeman}
		\author{J.J.~Krauth}
		\author{M.~Favier}
		\author{K.S.E.~Eikema}
		\affiliation{LaserLaB, Department of Physics and Astronomy, Vrije Universiteit, De Boelelaan 1081, 1081 HV Amsterdam, The Netherlands}
		\date{\today}
		
		\begin{abstract}
			High-harmonic generation (HHG) is widely used for up-conver\-sion of amplified (near) infrared ultrafast laser pulses to short wavelengths. We demonstrate that Ramsey-comb spectroscopy, based on two such pulses derived from a frequency-comb laser, enables us to observe phase effects in this process with a few mrad precision. As a result, we could perform the most accurate spectroscopic measurement based on light from HHG, illustrated with a determination of the $5p^6 \rightarrow 5p^5 8s~^2[3/2]_1$ transition at 110~nm in $^{132}$Xe. We improve its relative accuracy $10^4$ times to a value of $2.3\times10^{-10}$. This is 3.6 times better than shown before involving HHG, and promising to enable $1S-2S$ spectroscopy of He$^+$ for fundamental tests.
		\end{abstract}
		
		\maketitle
	\end{CJK*}

\begin{figure}[!t]
	\centering
	\includegraphics{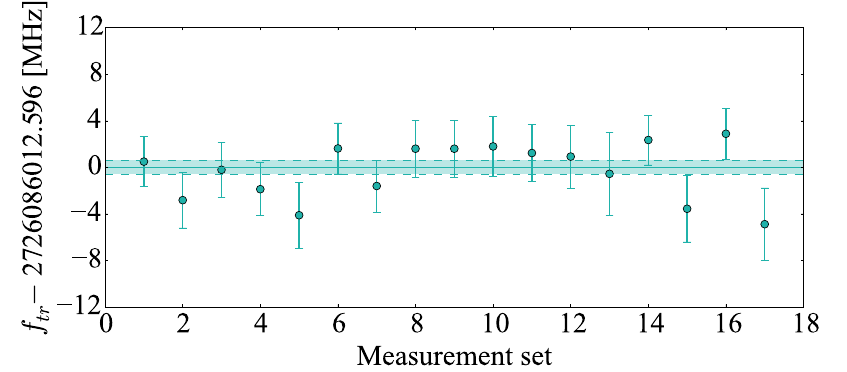}
	\caption{The Doppler-free transition frequency of the $5p^6\rightarrow 5p^58s~^2[3/2]_1$ transition in $^{132}$Xe at 110~nm. Each data point is based on 10 measurements at two different speeds of xenon. The blue shaded area indicates the 1$\sigma$ uncertainty.}
	\label{fig:transition_frequency_result} 
\end{figure}

High-precision spectroscopy in calculable atomic and molecular systems is at the heart of the most precise tests of bound-state quantum electrodynamics (QED) and searches for new physics beyond the Standard Model~\cite{parthey_improved_2011,holsch_benchmarking_2019,biesheuvel_probing_2016,stohlker_1_2000,kozlov_highly_2018,draginic_high_2003}. Instrumental in this development was the invention of the optical frequency comb (FC)~\cite{holzwarth_optical_2000,jones_carrier-envelope_2000} which enables precise optical frequency measurements referenced to an atomic clock. However, uncertainties in finite nuclear-size effects are hampering further progress~\cite{karshenboim_precision_2005}. Instead, spectroscopy has been used to measure the proton size in atomic and muonic hydrogen, but with partly conflicting results~\cite{pohl_size_2010,antognini_proton_2013,pohl_muonic_2013,antognini_experiments_2016,pohl_laser_2016,fleurbaey_new_2018,bezginov_measurement_2019}. 
High-precision spectroscopy of the $1S-2S$ transition in He$^+$ would provide new possibilities for fundamental tests as the uncertainty there is less dominated by nuclear size effects~\cite{herrmann_feasibility_2009}. Combined with muonic He$^+$ spectroscopy~\cite{franke_theory_2017,diepold_theory_2018} one can extract e.g.~the alpha particle radius or the Rydberg constant. 
A major experimental challenge arises from the requirement of  extreme ultraviolet (XUV) light at 60~nm (or shorter), to excite the transition. A similar challenge exist for spectroscopy of highly-charged ions~\cite{kozlov_highly_2018}, or the Thorium nuclear clock transition near 150~nm in the vacuum ultraviolet (VUV)~\cite{von_der_wense_direct_2016,von_der_wense_laser_2017}.
At those wavelengths a relative accuracy of 0.1 ppm has been achieved with Fourier-transform spectroscopy techniques~\cite{oliveira_high-resolution_2011}, and 0.03 ppm with low harmonics from nanosecond pulsed lasers~\cite{eikema_precision_1996}. A higher accuracy can be reached with light from high-harmonic generation (HHG), induced by focusing ultra-fast high-energy laser pulses in a noble gas at intensities of $\sim 10^{14}$~W/cm$^2$. The process can be understood using the three-step model~\cite{corkum_plasma_1993, lewenstein_theory_1994}, involving tunnel-ionization and recollision of an electron. This highly coherent process leads to the generation of a series of odd harmonics, which are tightly linked to the fundamental wave~\cite{zerne_phase-locked_1997,bellini_temporal_1998,gohle_frequency_2005,jones_phase-coherent_2005,benko_extreme_2014}. In combination with frequency-comb lasers, it has been used to achieve a spectroscopic accuracy of about 1 ppb at VUV and XUV wavelengths~\cite{kandula_extreme_2010,cingoz_direct_2012}.\\
To improve on this we recently developed the Ramsey-comb spectroscopy (RCS) method~\cite{morgenweg_ramsey-comb_2014,morgenweg_ramsey-comb_theory_2014}, based on pairs of powerful amplified FC pulses in a Ramsey-type~\cite{ramsey_molecular_1950} excitation scheme. Using only two pulses can compromise the accuracy provided by the FC laser~\cite{kandula_extreme_2010}, but the differential nature of RCS leads to the recovery of this accuracy~\cite{morgenweg_ramsey-comb_2014} and also to a strong suppression of the influence of the ac-Stark (light) shift. Other advantages of RCS compared to spectroscopy using cavity-based FC up-conversion~\cite{gohle_frequency_2005,jones_phase-coherent_2005,benko_extreme_2014} include easy tunability, simple up-conversion in a gas jet (no resonator required), a high excitation probability and a nearly 100$\%$ detection efficiency. RCS has been demonstrated successfully at wavelengths ranging from the near-infrared (NIR)~\cite{morgenweg_ramsey-comb_2014}, to the deep-ultraviolet, using low-order nonlinear up-conversion in crystals~\cite{altmann_high-precision_2016, altmann_deep-ultraviolet_2018}.\\
Extending RCS to shorter wavelengths with HHG is not trivial because a dynamic plasma is produced in HHG. This can lead to a reduced HHG yield and a time-dependent influence on the phase (and phase-matching) of the generated harmonics~\cite{corsi_direct_2006,rundquist_phase-matched_1998,popmintchev_phase_2009}, and therefore errors in the extracted transition frequency. Plasma effects in HHG have been investigated at picosecond-timescales, showing a nearly instantaneous response based on electron dynamics \cite{salieres_frequency-domain_1999,chen_measurement_2007}, but not at longer timescales relevant for precision spectroscopy. \\
In this Letter we show that the phase evolution of an atomic excitation obtained from RCS can be used to investigate the phase influence of plasma formation in HHG. We can monitor this on a nanosecond timescale and with mrad phase sensitivity. The results obtained at 110~nm (the seventh harmonic of 770~nm) show that most of the phase effects are caused by free electrons in the plasma and therefore strongly decrease within a few nanoseconds. Under the right conditions, the effects can be made negligibly small, enabling spectroscopy with unprecedented accuracy using radiation from HHG. This is demonstrated with a measurement of the $5p^6\rightarrow5p^58s~^2[3/2]_1$ transition in xenon with a relative accuracy of $2.3\times10^{-10}$.\\
\begin{figure*}
	\includegraphics{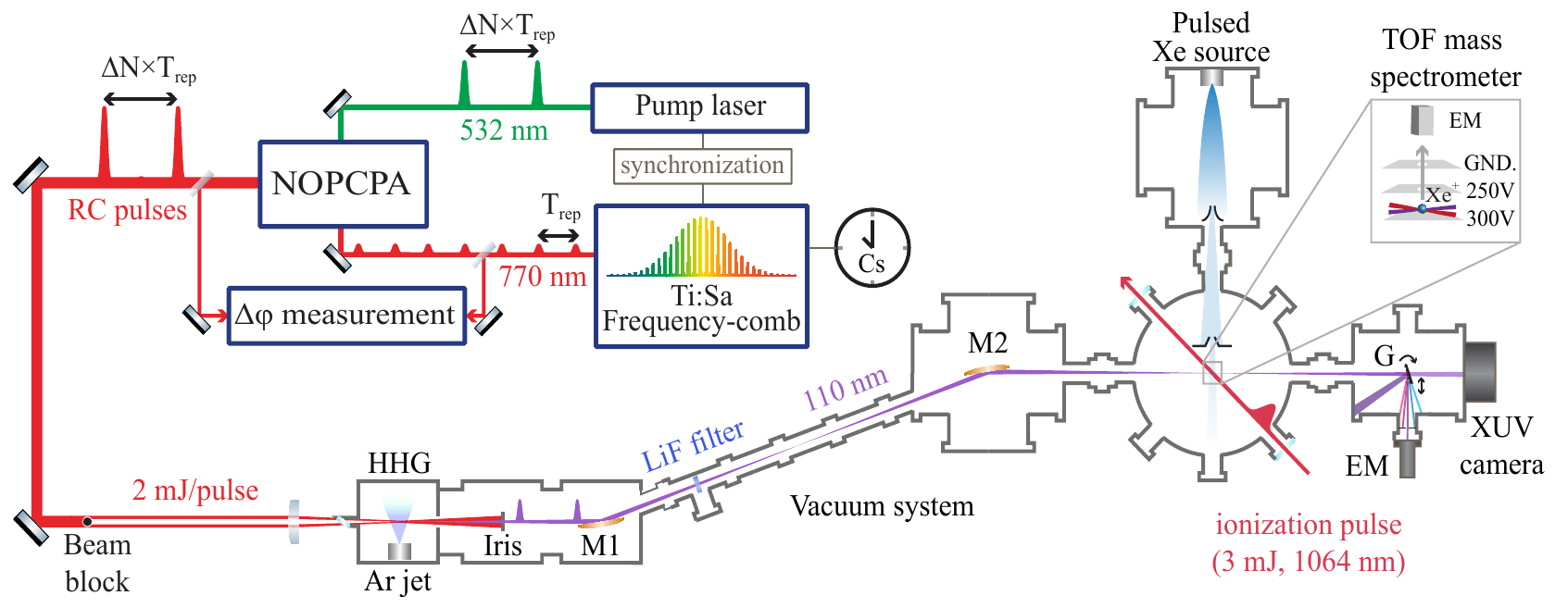}
	\caption{Schematic of the HHG-RCS setup. A pair of pulses from a Ti:Sa FC laser is selectively amplified at different multiples of the repetition time ($\Delta N \times T_{rep}$) in a NOPCPA. They are up-converted using HHG in an argon jet (shown 90$^{\circ}$ rotated). The seventh harmonic is refocused using a pair of toroidal mirrors ($M1$ and $M2$) at grazing incidence. Xenon atoms from a pulsed source are excited at 90$^{\circ}$ angle to reduce the first-order Doppler effect. An ionization pulse at 1064~nm is delayed by 2~ns with respect to the last excitation pulse and selectively ionizes the excited atoms. After pulsed field extraction, the ions are detected with an ETP AF880 electron multiplier (EM) at the end of a 47~cm long time-of-flight (TOF) drift tube. $G$ = grating (3600~lines/mm).}
	\label{fig:vacuum_setup}
\end{figure*}
Ramsey-comb spectroscopy~\cite{morgenweg_ramsey-comb_2014,morgenweg_ramsey-comb_theory_2014} requires phase and time-controlled laser pulses. The output of a FC laser~\cite{holzwarth_optical_2000,jones_carrier-envelope_2000} is the ideal source for this as both the repetition time $T_{rep}$ and carrier-envelope phase evolution $\Delta\phi_{ce}$ of the pulses can be referenced to an atomic clock. For two pulses that are resonant with a two-level system at a transition frequency $f_{tr}$, each excitation pulse can be thought to induce a superposition of the ground and excited state. These contributions interfere depending on the phase evolution $2\pi f_{tr} \Delta t$ of the superposition state (where $\Delta t$ is the pulse delay) and the phase difference $\Delta\phi$ between the two laser pulses (which includes $\Delta\phi_{ce}$). The excited state population after the second laser pulse can be written as $\lvert c_{e}(\Delta t, \Delta \phi)\rvert^2 \propto \cos(2\pi f_{tr}\Delta t+\Delta \phi)$. When a scan of $T_{rep}$ is made on a femtosecond or attosecond timescale, the effect of the phase evolution on $\lvert c_e\rvert^2$ can be observed in the form of a Ramsey fringe. A series of these fringes can be obtained at different multiples of the repetition time ($\Delta N \times T _{rep}$) by selecting different pairs of pulses. The transition frequency is then extracted from the phase \textit{difference} between these Ramsey fringes~\cite{morgenweg_ramsey-comb_theory_2014}, which leads to a cancellation of any induced, but constant, phase shift. This includes the optical phase shift between the two pulses (e.g.~from amplification), but also the ac-Stark light shift of the energy levels for a constant pulse energy~\cite{altmann_deep-ultraviolet_2018}.\\
The starting point of the laser system is a mode-locked Ti:sapphire FC laser ($T_{rep}=7.9$~ns) which is referenced to a Cs atomic clock (Symmetricon CsIII 4310B). The pulses are spectrally filtered within a 4f-grating stretcher to a bandwidth of 8-10~nm centered around 770~nm to avoid excitation of neighboring transitions.\\ 
A Non-collinear Optical Parametric Chirped Pulse Amplifier (NOPCPA), based on 3 Beta-Barium Borate (BBO) crystals, is used to selectively amplify two of the FC pulses. The delay between these pulses is an integer multiple of $T_{rep}$ and depends on the settings of a home-built double-pulse pump laser at 532~nm~\cite{morgenweg_tailored_2012,morgenweg_multi-delay_2013}. A typical energy of 2~mJ/pulse is reached after recompression to a $\sim$220 fs pulse length.\\
HHG is performed (see Fig.~\ref{fig:vacuum_setup}) by focusing ($f=25$ cm) the beam (4~mm FWHM diameter) in an argon gas jet. A central beam block of 1 mm diameter is used to convert the intensity profile of the fundamental beam to a donut-like shape before HHG. This enables efficient separation of the fundamental and harmonic beam with an adjustable iris after HHG, because the harmonics travel on axis and with a much lower divergence than the fundamental beam. A LiF plate blocks harmonics with $\lambda<105$~nm, while the seventh harmonic at 110~nm is transmitted with an efficiency of 40$\%$. The beam is subsequently refocused using a pair of grazing-incidence gold-coated toroidal mirrors, acting as a 1:1 telescope. The refocused VUV beam is crossed at 90$^\circ$ with a pulsed supersonic beam of xenon (backing pressure 500 mbar, pulse length $\approx 40$ $\mu$s). Excited atoms are detected by state selective ionization with a 3 mJ pulse at 1064~nm, and the resulting ions are recorded isotope-selectively with a time-of-flight mass separator. Despite previous reports of Xe cluster formation in supersonic expansions~\cite{wormer_fluorescence_1989,ozawa_vuv_2013}, no evidence for it was found in our experiment, even after many tests with different skimmers and valves.\\
\begin{figure}[!t]
	\centering
	\includegraphics{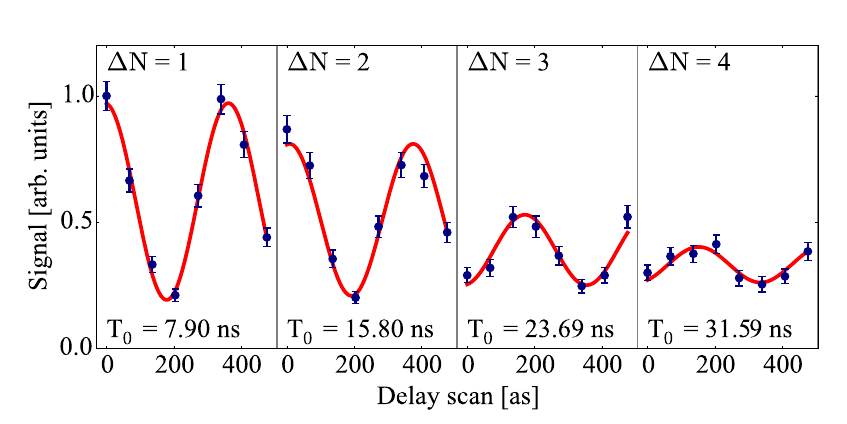}
	\caption{Typical Ramsey-comb scan of the $5p^6 \rightarrow 5p^5 8s~^2[3/2]_1$ transition at 110~nm in $^{132}$Xe. $\Delta N$ refers to the inter-pulse delay in multiples of $T_{rep}=7.9$~ns of the FC and $T_0$ indicates the initial delay. The individual data points are obtained by averaging over 700 laser shots, which leads to a measurement time of 3.3~min/fringe. The fringes are fitted with a fixed frequency to determine their phase.}
	\label{fig:Ramsey_fringes} 
\end{figure}
A typical RCS scan of $^{132}$Xe with $\Delta N = 1-4$ is shown in Fig.~\ref{fig:Ramsey_fringes}. Most measurements were done with only a pair of Ramsey fringes (e.g.~$\Delta N=2$ and $\Delta N=4$), taking just 6.6 minutes, to minimize the influence of possible drifts. For the same reason, the data points were recorded in random order and sorted according to pulse delay afterwards, instead of a sequential scan as was done in previous RCS experiments.\\
The contrast of the Ramsey fringes ($80-90$\% for $\Delta N=1$) decreases notably as a function of $\Delta N$. This is partly caused by the limited upper-state lifetime of $\approx22$~ns~\cite{chan_absolute_1992}, Doppler effects, and a 50-70~mrad rms phase noise of the amplified FC pulses. However, the decay of contrast was dominated by the limited transit time of the xenon atoms through the focused VUV beam. Therefore astigmatism was introduced to increase the interaction time from 16~ns (focus diameter 15 $\mu$m) to 32~ns (30 $\mu$m) at the expense of a maximum local wavefront tilt of $\approx 1.5$~mrad. This was inferred from the fundamental beam profile in the focal plane.\\ 
The combination of HHG with the refocusing optics revealed a subtle but interesting effect originating from the NOPCPA. An intensity (and alignment) dependent spatial walk-off induced effect led to a slight difference in beam pointing ($<$0.5~mrad) between the two amplified pulses. This reduced the overlap between the two refocused VUV pulses and further limited the interaction time with the atoms. It also led to a strong, initially unexplained, day-to-day variation of Ramsey signal contrast. After implementation of a walk-off compensating configuration~\cite{armstrong_parametric_1997} in the first two passes of the amplifier, the beam pointing difference was reduced to below 20 $\mu$rad. This was a crucial step to be able to combine HHG with RCS. In previous RCS experiments with low harmonics in crystals~\cite{morgenweg_ramsey-comb_2014,altmann_high-precision_2016,altmann_deep-ultraviolet_2018}, the walk-off induced beam pointing difference had little effect and went unnoticed because mm-size, collimated beams could be used.\\
\begin{figure}[!t]
	\centering
	\includegraphics{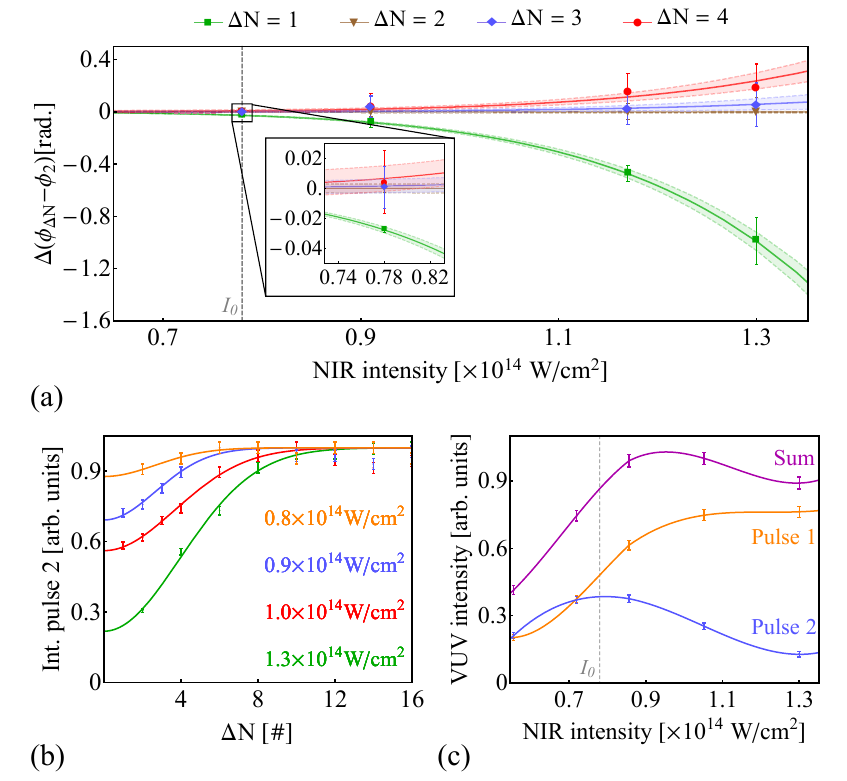}
	\caption{The influence of plasma-induced effects from HHG on the phase and the intensity of the second VUV pulse. (a) The phase difference between $\Delta N = 2$ (reference) and $\Delta N = 1,3$ and $4$ as a function of driving intensity. The data is fitted with $\phi\propto I^7$ and the shaded area indicates the $1\sigma$ uncertainty. The dashed line shows the intensity $I_0$ at normal operation and the inset shows a zoom-in at this value. (b) The intensity of the second VUV pulse as a function of delay for different driving intensities. (c) The harmonic yield of the two pulses and their sum as a function of driving intensity. The lines connecting the data points are splines to guide the eye.}
	\label{fig:HHG_intensity_phase} 
\end{figure}
With the aforementioned improvements, the influence of the generated plasma in the HHG process on the phase of the VUV light could be measured. For this we investigated the phase dependence of the Ramsey fringes on different conditions in the HHG process.\\
In Fig.~\ref{fig:HHG_intensity_phase}(a) the phase at $\Delta N = 1,3$ and $4$ relative to $\Delta N = 2$ as a function of the driving intensity is shown. This intensity was determined from the measured pulse energy, beam waist and pulse length (using frequency-resolved optical gating). We have chosen the phase at $\Delta N = 2$ as a reference, because the dynamics changes markedly at this delay (15.8~ns) and the signal quality was better than for the larger delays. The observed phase shift as a function of the intensity shows a near-exact seventh-order dependence (see fitted curves in Fig.~\ref{fig:HHG_intensity_phase}(a)). At $\Delta N = 1$, a maximum phase shift of 1 rad is observed. This is reduced by an order of magnitude at larger pulse delays, from which we conclude that the phase shift, especially at $\Delta N=1$, is dominated by the influence of fast moving free electrons (leaving the interaction zone on a ps to few ns time scale). As the atoms and ions move much slower, their contribution is seen mostly at later times.\\
This difference in dynamics between slow atoms (ions), and fast electrons is also illustrated in Fig.~\ref{fig:HHG_intensity_phase}(b), which shows the relative intensity of the second VUV pulse as a function of delay. The yield is significantly reduced up to 70$\%$ for high driving intensity because up-conversion of the first pulse leads to a significant depletion of neutral atoms. Note that this also leads to a reduction in contrast of the Ramsey fringes due to the imbalance between the two excitation contributions, and therefore a larger uncertainty for the phase (Fig.~\ref{fig:HHG_intensity_phase}(a)). The intensity of the second VUV pulse revives to a similar level as the first in 50-100~ns, depending on the driving intensity. This is in agreement with the expected transit time of argon atoms ($v \approx 500$~m/s) through the focus (50 $\mu$m) and it is much slower than the observed electron dynamics.\\
These results show that RCS can be combined successfully with HHG for precision measurements at short wavelengths. To demonstrate this we made an absolute calibration of the probed transition. Most of the observed phase effects occur at short pulse delays, and therefore only Ramsey fringes from $\Delta N = 2$ or higher were used to determine the transition frequency. In addition, the driving intensity was moderated to $I_0 = 0.78\times 10^{14}$~W/cm$^2$ (the dashed line in Fig.~\ref{fig:HHG_intensity_phase}(a) and (b)). The remaining shift in this case is $-2(5)$~mrad between $\Delta N = 2$ and $\Delta N = 3$ and $-7(9)$~mrad between $\Delta N = 2$ and $\Delta N = 4$ (inset in Fig.~\ref{fig:HHG_intensity_phase}(a)). This corresponds to a shift of $-32(91)$~kHz and $-67(86)$~kHz, respectively, of the extracted transition frequency and is consistent with zero within the uncertainty. Only a small penalty of 15-20\% of the maximum VUV yield (the sum of pulse 1 and 2) is paid by reducing the fundamental intensity (Fig.~\ref{fig:HHG_intensity_phase}(c)). The influence of the adiabatic HHG phase shift~\cite{lewenstein_theory_1994}, which depends on the driving NIR intensity, is suppressed in RCS, similar to ac-Stark shift. It is estimated to be below a few mrad ($<30$~kHz) in the VUV for an energy stability of $<0.2\%$ in the NIR.\\
The phase stability of the fundamental pulses is measured using spectral interferometry~\cite{morgenweg_multi-delay_2013} and found to be constant well within 1~mrad at 770~nm. The corresponding frequency uncertainty is $140$~kHz in the VUV.\\
The RCS signal is repetitive and leads to a frequency ambiguity of multiples of $f_{rep}=1/T_{rep}$~\cite{kandula_extreme_2010}. Therefore the measurements have been repeated with three slightly different values of $T_{rep}$ to obtain a single transition frequency with 99.2$\%$ confidence over a $4\sigma$ range of the former measurements~\cite{yoshino_absorption_1985}.\\
After the identification of the transition frequency, several systematic effects were investigated. The largest one was the Doppler effect, which was quantified and reduced by comparing the transition frequency obtained from pure xenon (velocity of 285(30)~m/s) with that from a 3:1 Ar:Xe mixture (480(30)~m/s). The angle between the atomic and the VUV beam was adjusted to minimize the observed frequency difference to a few MHz, after which the Doppler-free transition frequency was determined by extrapolation to zero velocity. In total, 300 measurements have been performed and the result is shown in Fig.~\ref{fig:transition_frequency_result}. Besides the Doppler effect, RCS is also affected by the recoil shift due to the absorption of a single photon \cite{cadoret_atom_2009}. This is accounted for by applying a correction of 125~kHz.\\
The RCS method strongly suppresses the influence of ac-Stark shifts~\cite{altmann_high-precision_2016,altmann_deep-ultraviolet_2018}. Measurements at different intensity levels were done to check for a residual effect, leading to an uncertainty of 20~kHz from the NIR light (estimated at 1$\times$10$^{11}$~W/cm$^2$) and $85$~kHz from the VUV intensity (estimated at 5$\times$10$^{7}$~W/cm$^2$). The DC-Stark shift and Zeeman shift were reduced by exciting in a field-compensated environment. An additional uncertainty of $20$~kHz and $52$~kHz, respectively, is taken into account for the possible influence of residual stray fields. \\
The final result of the $5p^6\rightarrow5p^58s~^2[3/2]_1$ transition frequency in $^{132}$Xe has a total accuracy of 630~kHz and the value together with the corrections and error budget is listed in Table~\ref{tab:error_budget_xe}.\\
To conclude, we demonstrated the first RCS measurement in combination with HHG. It enables extending RCS to much shorter wavelengths than what is possible with nonlinear crystals~\cite{altmann_high-precision_2016,altmann_deep-ultraviolet_2018}. However, HHG is also known to introduce detrimental phase shifts from plasma formation. These effects have been investigated with RCS using xenon atoms as a phase detector, leading to mrad-level sensitivity. We are able to discriminate between two effects which originate from different aspects of plasma formation. The intensity of the generated light is predominantly influenced by the depletion of neutral atoms. This effect persists for relatively long pulse delays (50-100~ns), because it depends on the dynamics of neutral atoms in a gas jet. On the other hand, the phase of the generated light is mainly affected by the dispersion from the generated free electrons. As these electrons move faster than the atoms, a significant reduction of this effect is already observed at $16$~ns, which enables RCS on the $5p^6\rightarrow5p^58s~^2[3/2]_1$ transition in xenon at 110~nm with a sub-MHz accuracy.\\ 
The obtained transition frequency is in good agreement with the previous determination~\cite{yoshino_absorption_1985}, but improves upon it by a factor $10^4$. With this measurement an unprecedented fractional uncertainty of $2.3\times 10^{-10}$ is achieved using light obtained with HHG.\\
The HHG process contributes remarkably little to the error budget of the frequency determination (Table~\ref{tab:error_budget_xe}). The only reason that previous RCS experiments~\cite{morgenweg_ramsey-comb_2014,altmann_high-precision_2016,altmann_deep-ultraviolet_2018} without HHG reached a higher accuracy was the longer pulse delay that could be applied in those cases. The short maximum transit time of 32~ns (and therefore pulse delay) in the current experiment is caused by the tight focus of the harmonic beam, because the setup has been designed for $1S-2S$ spectroscopy of a single trapped He$^+$ ion. This limited the accuracy for xenon as the influence of most systematic errors reduce proportionally to the pulse delay. Fortunately, future targets, such as He$^+$, can be trapped and observed for much longer times. Then pulse delays in the microsecond range can be used, with the added benefit that HHG phase effects are then effectively zero, even at high levels of ionization. This makes RCS very promising for precision measurements with (sub)kHz-level accuracy at VUV and XUV wavelengths.

	\begin{table}[h]
		\caption{Contributions (in kHz) to the measurement of the $5p^6\rightarrow5p^58s~^2[3/2]_1$ transition in xenon.}
		\label{tab:error_budget_xe}
		\begin{ruledtabular}
			\begin{tabular}{lrl}
				&Value or correction	&(1$\sigma$)	\\ 
				\hline 
				Doppler-free transition frequency	&$2\:726\:086\:012\:596$	&(600)\footnote{Including the uncertainty of $\approx$ 90 kHz due to the residual HHG phase shift (see text) and the correction for the second-order Doppler shift of 1.2 kHz for pure Xe and 3.5 kHz for the mixture.}			\\
				Light induced effects				&0				&(87)			\\
				DC-Stark shift						&0				&(20)			\\
				Zeeman-shift						&0				&(52)			\\
				Amplifier phase shift				&0				&(140)			\\ 
				Recoil shift						&-125			&($10^{-7}$)			\\  
				\hline 
				Total 								&$2\:726\:086\:012\:471$	&(630)			\\
			\end{tabular}
		\end{ruledtabular}
	\end{table}

	\begin{acknowledgments}
		We thank Prof. Dr. Fr\'{e}d\'{e}ric Merkt for valuable discussions during this experiment.\\ 
		K. E. acknowledges the European Research Council for ERC-Advanced grants under the European Union\textquotesingle s Horizon 2020 research and innovation programme (Grant Agreement No. 695677) and FOM/ NWO for a Program grant (16MYSTP).
	\end{acknowledgments}

	\bibliography{references}

\begin{thebibliography}{50}
\expandafter\ifx\csname natexlab\endcsname\relax\def\natexlab#1{#1}\fi
\expandafter\ifx\csname bibnamefont\endcsname\relax
  \def\bibnamefont#1{#1}\fi
\expandafter\ifx\csname bibfnamefont\endcsname\relax
  \def\bibfnamefont#1{#1}\fi
\expandafter\ifx\csname citenamefont\endcsname\relax
  \def\citenamefont#1{#1}\fi
\expandafter\ifx\csname url\endcsname\relax
  \def\url#1{\texttt{#1}}\fi
\expandafter\ifx\csname urlprefix\endcsname\relax\def\urlprefix{URL }\fi
\providecommand{\bibinfo}[2]{#2}
\providecommand{\eprint}[2][]{\url{#2}}

\bibitem[{\citenamefont{Parthey et~al.}(2011)\citenamefont{Parthey, Matveev,
  Alnis, Bernhardt, Beyer, Holzwarth, Maistrou, Pohl, Predehl, Udem
  et~al.}}]{parthey_improved_2011}
\bibinfo{author}{\bibfnamefont{C.~G.} \bibnamefont{Parthey}},
  \bibinfo{author}{\bibfnamefont{A.}~\bibnamefont{Matveev}},
  \bibinfo{author}{\bibfnamefont{J.}~\bibnamefont{Alnis}},
  \bibinfo{author}{\bibfnamefont{B.}~\bibnamefont{Bernhardt}},
  \bibinfo{author}{\bibfnamefont{A.}~\bibnamefont{Beyer}},
  \bibinfo{author}{\bibfnamefont{R.}~\bibnamefont{Holzwarth}},
  \bibinfo{author}{\bibfnamefont{A.}~\bibnamefont{Maistrou}},
  \bibinfo{author}{\bibfnamefont{R.}~\bibnamefont{Pohl}},
  \bibinfo{author}{\bibfnamefont{K.}~\bibnamefont{Predehl}},
  \bibinfo{author}{\bibfnamefont{T.}~\bibnamefont{Udem}}, \bibnamefont{et~al.},
  \bibinfo{journal}{Phys. Rev. Lett.} \textbf{\bibinfo{volume}{107}},
  \bibinfo{pages}{203001} (\bibinfo{year}{2011}).

\bibitem[{\citenamefont{H\"olsch et~al.}(2019)\citenamefont{H\"olsch, Beyer,
  Salumbides, Eikema, Ubachs, Jungen, and Merkt}}]{holsch_benchmarking_2019}
\bibinfo{author}{\bibfnamefont{N.}~\bibnamefont{H\"olsch}},
  \bibinfo{author}{\bibfnamefont{M.}~\bibnamefont{Beyer}},
  \bibinfo{author}{\bibfnamefont{E.~J.} \bibnamefont{Salumbides}},
  \bibinfo{author}{\bibfnamefont{K.~S.~E.} \bibnamefont{Eikema}},
  \bibinfo{author}{\bibfnamefont{W.}~\bibnamefont{Ubachs}},
  \bibinfo{author}{\bibfnamefont{C.}~\bibnamefont{Jungen}}, \bibnamefont{and}
  \bibinfo{author}{\bibfnamefont{F.}~\bibnamefont{Merkt}},
  \bibinfo{journal}{Phys. Rev. Lett.} \textbf{\bibinfo{volume}{122}},
  \bibinfo{pages}{103002} (\bibinfo{year}{2019}).

\bibitem[{\citenamefont{Biesheuvel et~al.}(2016)\citenamefont{Biesheuvel, Karr,
  Hilico, Eikema, Ubachs, and Koelemeij}}]{biesheuvel_probing_2016}
\bibinfo{author}{\bibfnamefont{J.}~\bibnamefont{Biesheuvel}},
  \bibinfo{author}{\bibfnamefont{J.-P.} \bibnamefont{Karr}},
  \bibinfo{author}{\bibfnamefont{L.}~\bibnamefont{Hilico}},
  \bibinfo{author}{\bibfnamefont{K.~S.~E.} \bibnamefont{Eikema}},
  \bibinfo{author}{\bibfnamefont{W.}~\bibnamefont{Ubachs}}, \bibnamefont{and}
  \bibinfo{author}{\bibfnamefont{J.~C.~J.} \bibnamefont{Koelemeij}},
  \bibinfo{journal}{Nat. Commun.} \textbf{\bibinfo{volume}{7}}
  (\bibinfo{year}{2016}).

\bibitem[{\citenamefont{St\"ohlker et~al.}(2000)\citenamefont{St\"ohlker,
  Mokler, Bosch, Dunford, Franzke, Klepper, Kozhuharov, Ludziejewski, Nolden,
  Reich et~al.}}]{stohlker_1_2000}
\bibinfo{author}{\bibfnamefont{T.}~\bibnamefont{St\"ohlker}},
  \bibinfo{author}{\bibfnamefont{P.~H.} \bibnamefont{Mokler}},
  \bibinfo{author}{\bibfnamefont{F.}~\bibnamefont{Bosch}},
  \bibinfo{author}{\bibfnamefont{R.~W.} \bibnamefont{Dunford}},
  \bibinfo{author}{\bibfnamefont{F.}~\bibnamefont{Franzke}},
  \bibinfo{author}{\bibfnamefont{O.}~\bibnamefont{Klepper}},
  \bibinfo{author}{\bibfnamefont{C.}~\bibnamefont{Kozhuharov}},
  \bibinfo{author}{\bibfnamefont{T.}~\bibnamefont{Ludziejewski}},
  \bibinfo{author}{\bibfnamefont{F.}~\bibnamefont{Nolden}},
  \bibinfo{author}{\bibfnamefont{H.}~\bibnamefont{Reich}},
  \bibnamefont{et~al.}, \bibinfo{journal}{Phys. Rev. Lett.}
  \textbf{\bibinfo{volume}{85}}, \bibinfo{pages}{3109} (\bibinfo{year}{2000}).

\bibitem[{\citenamefont{Kozlov et~al.}(2018)\citenamefont{Kozlov, Safronova,
  Crespo L\'opez-Urrutia, and Schmidt}}]{kozlov_highly_2018}
\bibinfo{author}{\bibfnamefont{M.~G.} \bibnamefont{Kozlov}},
  \bibinfo{author}{\bibfnamefont{M.~S.} \bibnamefont{Safronova}},
  \bibinfo{author}{\bibfnamefont{J.~R.} \bibnamefont{Crespo L\'opez-Urrutia}},
  \bibnamefont{and} \bibinfo{author}{\bibfnamefont{P.~O.}
  \bibnamefont{Schmidt}}, \bibinfo{journal}{Rev. Mod. Phys.}
  \textbf{\bibinfo{volume}{90}}, \bibinfo{pages}{045005}
  (\bibinfo{year}{2018}).

\bibitem[{\citenamefont{Dragani\ifmmode~\acute{c}\else \'{c}\fi{}
  et~al.}(2003)\citenamefont{Dragani\ifmmode~\acute{c}\else \'{c}\fi{}, Crespo
  L\'opez-Urrutia, DuBois, Fritzsche, Shabaev, Orts, Tupitsyn, Zou, and
  Ullrich}}]{draginic_high_2003}
\bibinfo{author}{\bibfnamefont{I.}~\bibnamefont{Dragani\ifmmode~\acute{c}\else
  \'{c}\fi{}}}, \bibinfo{author}{\bibfnamefont{J.~R.} \bibnamefont{Crespo
  L\'opez-Urrutia}}, \bibinfo{author}{\bibfnamefont{R.}~\bibnamefont{DuBois}},
  \bibinfo{author}{\bibfnamefont{S.}~\bibnamefont{Fritzsche}},
  \bibinfo{author}{\bibfnamefont{V.~M.} \bibnamefont{Shabaev}},
  \bibinfo{author}{\bibfnamefont{R.~S.} \bibnamefont{Orts}},
  \bibinfo{author}{\bibfnamefont{I.~I.} \bibnamefont{Tupitsyn}},
  \bibinfo{author}{\bibfnamefont{Y.}~\bibnamefont{Zou}}, \bibnamefont{and}
  \bibinfo{author}{\bibfnamefont{J.}~\bibnamefont{Ullrich}},
  \bibinfo{journal}{Phys. Rev. Lett.} \textbf{\bibinfo{volume}{91}},
  \bibinfo{pages}{183001} (\bibinfo{year}{2003}).

\bibitem[{\citenamefont{Holzwarth et~al.}(2000)\citenamefont{Holzwarth, Udem,
  H\"ansch, Knight, Wadsworth, and Russell}}]{holzwarth_optical_2000}
\bibinfo{author}{\bibfnamefont{R.}~\bibnamefont{Holzwarth}},
  \bibinfo{author}{\bibfnamefont{T.}~\bibnamefont{Udem}},
  \bibinfo{author}{\bibfnamefont{T.~W.} \bibnamefont{H\"ansch}},
  \bibinfo{author}{\bibfnamefont{J.~C.} \bibnamefont{Knight}},
  \bibinfo{author}{\bibfnamefont{W.~J.} \bibnamefont{Wadsworth}},
  \bibnamefont{and} \bibinfo{author}{\bibfnamefont{P.~S.~J.}
  \bibnamefont{Russell}}, \bibinfo{journal}{Phys. Rev. Lett.}
  \textbf{\bibinfo{volume}{85}}, \bibinfo{pages}{2264} (\bibinfo{year}{2000}).

\bibitem[{\citenamefont{Jones et~al.}(2000)\citenamefont{Jones, Diddams, Ranka,
  Stentz, Windeler, Hall, and Cundiff}}]{jones_carrier-envelope_2000}
\bibinfo{author}{\bibfnamefont{D.~J.} \bibnamefont{Jones}},
  \bibinfo{author}{\bibfnamefont{S.~A.} \bibnamefont{Diddams}},
  \bibinfo{author}{\bibfnamefont{J.~K.} \bibnamefont{Ranka}},
  \bibinfo{author}{\bibfnamefont{A.}~\bibnamefont{Stentz}},
  \bibinfo{author}{\bibfnamefont{R.~S.} \bibnamefont{Windeler}},
  \bibinfo{author}{\bibfnamefont{J.~L.} \bibnamefont{Hall}}, \bibnamefont{and}
  \bibinfo{author}{\bibfnamefont{S.~T.} \bibnamefont{Cundiff}},
  \bibinfo{journal}{Science} \textbf{\bibinfo{volume}{288}},
  \bibinfo{pages}{635} (\bibinfo{year}{2000}).

\bibitem[{\citenamefont{Karshenboim}(2005)}]{karshenboim_precision_2005}
\bibinfo{author}{\bibfnamefont{S.~G.} \bibnamefont{Karshenboim}},
  \bibinfo{journal}{Phys. Rep.} \textbf{\bibinfo{volume}{422}},
  \bibinfo{pages}{1} (\bibinfo{year}{2005}).

\bibitem[{\citenamefont{Pohl et~al.}(2010)\citenamefont{Pohl, Antognini, Nez,
  Amaro, Biraben, Cardoso, Covita, Dax, Dhawan, Fernandes
  et~al.}}]{pohl_size_2010}
\bibinfo{author}{\bibfnamefont{R.}~\bibnamefont{Pohl}},
  \bibinfo{author}{\bibfnamefont{A.}~\bibnamefont{Antognini}},
  \bibinfo{author}{\bibfnamefont{F.}~\bibnamefont{Nez}},
  \bibinfo{author}{\bibfnamefont{F.~D.} \bibnamefont{Amaro}},
  \bibinfo{author}{\bibfnamefont{F.}~\bibnamefont{Biraben}},
  \bibinfo{author}{\bibfnamefont{J.~a. M.~R.} \bibnamefont{Cardoso}},
  \bibinfo{author}{\bibfnamefont{D.~S.} \bibnamefont{Covita}},
  \bibinfo{author}{\bibfnamefont{A.}~\bibnamefont{Dax}},
  \bibinfo{author}{\bibfnamefont{S.}~\bibnamefont{Dhawan}},
  \bibinfo{author}{\bibfnamefont{L.~M.~P.} \bibnamefont{Fernandes}},
  \bibnamefont{et~al.}, \bibinfo{journal}{Nature (London)}
  \textbf{\bibinfo{volume}{466}}, \bibinfo{pages}{213} (\bibinfo{year}{2010}).

\bibitem[{\citenamefont{Antognini et~al.}(2013)\citenamefont{Antognini, Nez,
  Schuhmann, Amaro, Biraben, Cardoso, Covita, Dax, Dhawan, Diepold
  et~al.}}]{antognini_proton_2013}
\bibinfo{author}{\bibfnamefont{A.}~\bibnamefont{Antognini}},
  \bibinfo{author}{\bibfnamefont{F.}~\bibnamefont{Nez}},
  \bibinfo{author}{\bibfnamefont{K.}~\bibnamefont{Schuhmann}},
  \bibinfo{author}{\bibfnamefont{F.~D.} \bibnamefont{Amaro}},
  \bibinfo{author}{\bibfnamefont{F.}~\bibnamefont{Biraben}},
  \bibinfo{author}{\bibfnamefont{J.~M.~R.} \bibnamefont{Cardoso}},
  \bibinfo{author}{\bibfnamefont{D.~S.} \bibnamefont{Covita}},
  \bibinfo{author}{\bibfnamefont{A.}~\bibnamefont{Dax}},
  \bibinfo{author}{\bibfnamefont{S.}~\bibnamefont{Dhawan}},
  \bibinfo{author}{\bibfnamefont{M.}~\bibnamefont{Diepold}},
  \bibnamefont{et~al.}, \bibinfo{journal}{Science}
  \textbf{\bibinfo{volume}{339}}, \bibinfo{pages}{417} (\bibinfo{year}{2013}).

\bibitem[{\citenamefont{Pohl et~al.}(2013)\citenamefont{Pohl, Gilman, Miller,
  and Pachucki}}]{pohl_muonic_2013}
\bibinfo{author}{\bibfnamefont{R.}~\bibnamefont{Pohl}},
  \bibinfo{author}{\bibfnamefont{R.}~\bibnamefont{Gilman}},
  \bibinfo{author}{\bibfnamefont{G.~A.} \bibnamefont{Miller}},
  \bibnamefont{and} \bibinfo{author}{\bibfnamefont{K.}~\bibnamefont{Pachucki}},
  \bibinfo{journal}{Annu. Rev. Nucl. Part. Sci.} \textbf{\bibinfo{volume}{63}},
  \bibinfo{pages}{175} (\bibinfo{year}{2013}).

\bibitem[{\citenamefont{Antognini et~al.}(2016)\citenamefont{Antognini,
  Schuhmann, Amaro, Amaro, {Abdou-Ahmed}, Biraben, Chen, Covita, Dax, Diepold
  et~al.}}]{antognini_experiments_2016}
\bibinfo{author}{\bibfnamefont{A.}~\bibnamefont{Antognini}},
  \bibinfo{author}{\bibfnamefont{K.}~\bibnamefont{Schuhmann}},
  \bibinfo{author}{\bibfnamefont{F.~D.} \bibnamefont{Amaro}},
  \bibinfo{author}{\bibfnamefont{P.}~\bibnamefont{Amaro}},
  \bibinfo{author}{\bibfnamefont{M.}~\bibnamefont{{Abdou-Ahmed}}},
  \bibinfo{author}{\bibfnamefont{F.}~\bibnamefont{Biraben}},
  \bibinfo{author}{\bibfnamefont{T.-L.} \bibnamefont{Chen}},
  \bibinfo{author}{\bibfnamefont{D.~S.} \bibnamefont{Covita}},
  \bibinfo{author}{\bibfnamefont{A.~J.} \bibnamefont{Dax}},
  \bibinfo{author}{\bibfnamefont{M.}~\bibnamefont{Diepold}},
  \bibnamefont{et~al.}, \bibinfo{journal}{EPJ Web Conf.}
  \textbf{\bibinfo{volume}{113}}, \bibinfo{pages}{01006}
  (\bibinfo{year}{2016}).

\bibitem[{\citenamefont{Pohl et~al.}(2016)\citenamefont{Pohl, Nez, Fernandes,
  Amaro, Biraben, Cardoso, Covita, Dax, Dhawan, Diepold
  et~al.}}]{pohl_laser_2016}
\bibinfo{author}{\bibfnamefont{R.}~\bibnamefont{Pohl}},
  \bibinfo{author}{\bibfnamefont{F.}~\bibnamefont{Nez}},
  \bibinfo{author}{\bibfnamefont{L.~M.~P.} \bibnamefont{Fernandes}},
  \bibinfo{author}{\bibfnamefont{F.~D.} \bibnamefont{Amaro}},
  \bibinfo{author}{\bibfnamefont{F.}~\bibnamefont{Biraben}},
  \bibinfo{author}{\bibfnamefont{J.~M.~R.} \bibnamefont{Cardoso}},
  \bibinfo{author}{\bibfnamefont{D.~S.} \bibnamefont{Covita}},
  \bibinfo{author}{\bibfnamefont{A.}~\bibnamefont{Dax}},
  \bibinfo{author}{\bibfnamefont{S.}~\bibnamefont{Dhawan}},
  \bibinfo{author}{\bibfnamefont{M.}~\bibnamefont{Diepold}},
  \bibnamefont{et~al.}, \bibinfo{journal}{Science}
  \textbf{\bibinfo{volume}{353}}, \bibinfo{pages}{669} (\bibinfo{year}{2016}).

\bibitem[{\citenamefont{Fleurbaey et~al.}(2018)\citenamefont{Fleurbaey,
  Galtier, Thomas, Bonnaud, Julien, Biraben, Nez, Abgrall, and
  Gu\'ena}}]{fleurbaey_new_2018}
\bibinfo{author}{\bibfnamefont{H.}~\bibnamefont{Fleurbaey}},
  \bibinfo{author}{\bibfnamefont{S.}~\bibnamefont{Galtier}},
  \bibinfo{author}{\bibfnamefont{S.}~\bibnamefont{Thomas}},
  \bibinfo{author}{\bibfnamefont{M.}~\bibnamefont{Bonnaud}},
  \bibinfo{author}{\bibfnamefont{L.}~\bibnamefont{Julien}},
  \bibinfo{author}{\bibfnamefont{F.}~\bibnamefont{Biraben}},
  \bibinfo{author}{\bibfnamefont{F.}~\bibnamefont{Nez}},
  \bibinfo{author}{\bibfnamefont{M.}~\bibnamefont{Abgrall}}, \bibnamefont{and}
  \bibinfo{author}{\bibfnamefont{J.}~\bibnamefont{Gu\'ena}},
  \bibinfo{journal}{Phys. Rev. Lett.} \textbf{\bibinfo{volume}{120}},
  \bibinfo{pages}{183001} (\bibinfo{year}{2018}).

\bibitem[{\citenamefont{Bezginov et~al.}(2019)\citenamefont{Bezginov, Valdez,
  Horbatsch, Marsman, Vutha, and Hessels}}]{bezginov_measurement_2019}
\bibinfo{author}{\bibfnamefont{N.}~\bibnamefont{Bezginov}},
  \bibinfo{author}{\bibfnamefont{T.}~\bibnamefont{Valdez}},
  \bibinfo{author}{\bibfnamefont{M.}~\bibnamefont{Horbatsch}},
  \bibinfo{author}{\bibfnamefont{A.}~\bibnamefont{Marsman}},
  \bibinfo{author}{\bibfnamefont{A.~C.} \bibnamefont{Vutha}}, \bibnamefont{and}
  \bibinfo{author}{\bibfnamefont{E.~A.} \bibnamefont{Hessels}},
  \bibinfo{journal}{Science} \textbf{\bibinfo{volume}{365}},
  \bibinfo{pages}{1007} (\bibinfo{year}{2019}).

\bibitem[{\citenamefont{Herrmann et~al.}(2009)\citenamefont{Herrmann, Haas,
  Jentschura, Kottmann, Leibfried, Saathoff, Gohle, Ozawa, Batteiger, Kn\"unz
  et~al.}}]{herrmann_feasibility_2009}
\bibinfo{author}{\bibfnamefont{M.}~\bibnamefont{Herrmann}},
  \bibinfo{author}{\bibfnamefont{M.}~\bibnamefont{Haas}},
  \bibinfo{author}{\bibfnamefont{U.~D.} \bibnamefont{Jentschura}},
  \bibinfo{author}{\bibfnamefont{F.}~\bibnamefont{Kottmann}},
  \bibinfo{author}{\bibfnamefont{D.}~\bibnamefont{Leibfried}},
  \bibinfo{author}{\bibfnamefont{G.}~\bibnamefont{Saathoff}},
  \bibinfo{author}{\bibfnamefont{C.}~\bibnamefont{Gohle}},
  \bibinfo{author}{\bibfnamefont{A.}~\bibnamefont{Ozawa}},
  \bibinfo{author}{\bibfnamefont{V.}~\bibnamefont{Batteiger}},
  \bibinfo{author}{\bibfnamefont{S.}~\bibnamefont{Kn\"unz}},
  \bibnamefont{et~al.}, \bibinfo{journal}{Phys. Rev. A}
  \textbf{\bibinfo{volume}{79}}, \bibinfo{pages}{052505}
  (\bibinfo{year}{2009}).

\bibitem[{\citenamefont{Franke et~al.}(2017)\citenamefont{Franke, Krauth,
  Antognini, Diepold, Kottmann, and Pohl}}]{franke_theory_2017}
\bibinfo{author}{\bibfnamefont{B.}~\bibnamefont{Franke}},
  \bibinfo{author}{\bibfnamefont{J.~J.} \bibnamefont{Krauth}},
  \bibinfo{author}{\bibfnamefont{A.}~\bibnamefont{Antognini}},
  \bibinfo{author}{\bibfnamefont{M.}~\bibnamefont{Diepold}},
  \bibinfo{author}{\bibfnamefont{F.}~\bibnamefont{Kottmann}}, \bibnamefont{and}
  \bibinfo{author}{\bibfnamefont{R.}~\bibnamefont{Pohl}},
  \bibinfo{journal}{Eur. Phys. J. D} \textbf{\bibinfo{volume}{71}},
  \bibinfo{pages}{341} (\bibinfo{year}{2017}).

\bibitem[{\citenamefont{Diepold et~al.}(2018)\citenamefont{Diepold, Franke,
  Krauth, Antognini, Kottmann, and Pohl}}]{diepold_theory_2018}
\bibinfo{author}{\bibfnamefont{M.}~\bibnamefont{Diepold}},
  \bibinfo{author}{\bibfnamefont{B.}~\bibnamefont{Franke}},
  \bibinfo{author}{\bibfnamefont{J.~J.} \bibnamefont{Krauth}},
  \bibinfo{author}{\bibfnamefont{A.}~\bibnamefont{Antognini}},
  \bibinfo{author}{\bibfnamefont{F.}~\bibnamefont{Kottmann}}, \bibnamefont{and}
  \bibinfo{author}{\bibfnamefont{R.}~\bibnamefont{Pohl}},
  \bibinfo{journal}{Ann. Phys.} \textbf{\bibinfo{volume}{396}},
  \bibinfo{pages}{220 } (\bibinfo{year}{2018}).

\bibitem[{\citenamefont{{von der Wense} et~al.}(2016)\citenamefont{{von der
  Wense}, Seiferle, Laatiaoui, Neumayr, Maier, Wirth, Mokry, Runke, Eberhardt,
  D{\"u}llmann et~al.}}]{von_der_wense_direct_2016}
\bibinfo{author}{\bibfnamefont{L.}~\bibnamefont{{von der Wense}}},
  \bibinfo{author}{\bibfnamefont{B.}~\bibnamefont{Seiferle}},
  \bibinfo{author}{\bibfnamefont{M.}~\bibnamefont{Laatiaoui}},
  \bibinfo{author}{\bibfnamefont{J.~B.} \bibnamefont{Neumayr}},
  \bibinfo{author}{\bibfnamefont{H.-J.} \bibnamefont{Maier}},
  \bibinfo{author}{\bibfnamefont{H.-F.} \bibnamefont{Wirth}},
  \bibinfo{author}{\bibfnamefont{C.}~\bibnamefont{Mokry}},
  \bibinfo{author}{\bibfnamefont{J.}~\bibnamefont{Runke}},
  \bibinfo{author}{\bibfnamefont{K.}~\bibnamefont{Eberhardt}},
  \bibinfo{author}{\bibfnamefont{C.~E.} \bibnamefont{D{\"u}llmann}},
  \bibnamefont{et~al.}, \bibinfo{journal}{Nature (London)}
  \textbf{\bibinfo{volume}{533}}, \bibinfo{pages}{47} (\bibinfo{year}{2016}).

\bibitem[{\citenamefont{von~der Wense et~al.}(2017)\citenamefont{von~der Wense,
  Seiferle, Stellmer, Weitenberg, Kazakov, P\'alffy, and
  Thirolf}}]{von_der_wense_laser_2017}
\bibinfo{author}{\bibfnamefont{L.}~\bibnamefont{von~der Wense}},
  \bibinfo{author}{\bibfnamefont{B.}~\bibnamefont{Seiferle}},
  \bibinfo{author}{\bibfnamefont{S.}~\bibnamefont{Stellmer}},
  \bibinfo{author}{\bibfnamefont{J.}~\bibnamefont{Weitenberg}},
  \bibinfo{author}{\bibfnamefont{G.}~\bibnamefont{Kazakov}},
  \bibinfo{author}{\bibfnamefont{A.}~\bibnamefont{P\'alffy}}, \bibnamefont{and}
  \bibinfo{author}{\bibfnamefont{P.~G.} \bibnamefont{Thirolf}},
  \bibinfo{journal}{Phys. Rev. Lett.} \textbf{\bibinfo{volume}{119}},
  \bibinfo{pages}{132503} (\bibinfo{year}{2017}).

\bibitem[{\citenamefont{de~Oliveira et~al.}(2011)\citenamefont{de~Oliveira,
  Roudjane, Joyeux, Phalippou, Rodier, and
  Nahon}}]{oliveira_high-resolution_2011}
\bibinfo{author}{\bibfnamefont{N.}~\bibnamefont{de~Oliveira}},
  \bibinfo{author}{\bibfnamefont{M.}~\bibnamefont{Roudjane}},
  \bibinfo{author}{\bibfnamefont{D.}~\bibnamefont{Joyeux}},
  \bibinfo{author}{\bibfnamefont{D.}~\bibnamefont{Phalippou}},
  \bibinfo{author}{\bibfnamefont{J.-C.} \bibnamefont{Rodier}},
  \bibnamefont{and} \bibinfo{author}{\bibfnamefont{L.}~\bibnamefont{Nahon}},
  \bibinfo{journal}{Nat. Photonics} \textbf{\bibinfo{volume}{5}},
  \bibinfo{pages}{149 EP } (\bibinfo{year}{2011}).

\bibitem[{\citenamefont{Eikema et~al.}(1996)\citenamefont{Eikema, Ubachs,
  Vassen, and Hogervorst}}]{eikema_precision_1996}
\bibinfo{author}{\bibfnamefont{K.~S.~E.} \bibnamefont{Eikema}},
  \bibinfo{author}{\bibfnamefont{W.}~\bibnamefont{Ubachs}},
  \bibinfo{author}{\bibfnamefont{W.}~\bibnamefont{Vassen}}, \bibnamefont{and}
  \bibinfo{author}{\bibfnamefont{W.}~\bibnamefont{Hogervorst}},
  \bibinfo{journal}{Phys. Rev. Lett.} \textbf{\bibinfo{volume}{76}},
  \bibinfo{pages}{1216} (\bibinfo{year}{1996}).

\bibitem[{\citenamefont{Corkum}(1993)}]{corkum_plasma_1993}
\bibinfo{author}{\bibfnamefont{P.~B.} \bibnamefont{Corkum}},
  \bibinfo{journal}{Phys. Rev. Lett.} \textbf{\bibinfo{volume}{71}},
  \bibinfo{pages}{1994} (\bibinfo{year}{1993}).

\bibitem[{\citenamefont{Lewenstein et~al.}(1994)\citenamefont{Lewenstein,
  Balcou, Ivanov, L'Huillier, and Corkum}}]{lewenstein_theory_1994}
\bibinfo{author}{\bibfnamefont{M.}~\bibnamefont{Lewenstein}},
  \bibinfo{author}{\bibfnamefont{P.}~\bibnamefont{Balcou}},
  \bibinfo{author}{\bibfnamefont{M.~Y.} \bibnamefont{Ivanov}},
  \bibinfo{author}{\bibfnamefont{A.}~\bibnamefont{L'Huillier}},
  \bibnamefont{and} \bibinfo{author}{\bibfnamefont{P.~B.}
  \bibnamefont{Corkum}}, \bibinfo{journal}{Phys. Rev. A}
  \textbf{\bibinfo{volume}{49}}, \bibinfo{pages}{2117} (\bibinfo{year}{1994}).

\bibitem[{\citenamefont{Zerne et~al.}(1997)\citenamefont{Zerne, Altucci,
  Bellini, Gaarde, H\"ansch, L'Huillier, Lyng\aa, and
  Wahlstr\"om}}]{zerne_phase-locked_1997}
\bibinfo{author}{\bibfnamefont{R.}~\bibnamefont{Zerne}},
  \bibinfo{author}{\bibfnamefont{C.}~\bibnamefont{Altucci}},
  \bibinfo{author}{\bibfnamefont{M.}~\bibnamefont{Bellini}},
  \bibinfo{author}{\bibfnamefont{M.~B.} \bibnamefont{Gaarde}},
  \bibinfo{author}{\bibfnamefont{T.~W.} \bibnamefont{H\"ansch}},
  \bibinfo{author}{\bibfnamefont{A.}~\bibnamefont{L'Huillier}},
  \bibinfo{author}{\bibfnamefont{C.}~\bibnamefont{Lyng\aa}}, \bibnamefont{and}
  \bibinfo{author}{\bibfnamefont{C.-G.} \bibnamefont{Wahlstr\"om}},
  \bibinfo{journal}{Phys. Rev. Lett.} \textbf{\bibinfo{volume}{79}},
  \bibinfo{pages}{1006} (\bibinfo{year}{1997}).

\bibitem[{\citenamefont{Bellini et~al.}(1998)\citenamefont{Bellini, Lyng\aa{},
  Tozzi, Gaarde, H\"ansch, L'Huillier, and
  Wahlstr\"om}}]{bellini_temporal_1998}
\bibinfo{author}{\bibfnamefont{M.}~\bibnamefont{Bellini}},
  \bibinfo{author}{\bibfnamefont{C.}~\bibnamefont{Lyng\aa{}}},
  \bibinfo{author}{\bibfnamefont{A.}~\bibnamefont{Tozzi}},
  \bibinfo{author}{\bibfnamefont{M.~B.} \bibnamefont{Gaarde}},
  \bibinfo{author}{\bibfnamefont{T.~W.} \bibnamefont{H\"ansch}},
  \bibinfo{author}{\bibfnamefont{A.}~\bibnamefont{L'Huillier}},
  \bibnamefont{and} \bibinfo{author}{\bibfnamefont{C.-G.}
  \bibnamefont{Wahlstr\"om}}, \bibinfo{journal}{Phys. Rev. Lett.}
  \textbf{\bibinfo{volume}{81}}, \bibinfo{pages}{297} (\bibinfo{year}{1998}).

\bibitem[{\citenamefont{Gohle et~al.}(2005)\citenamefont{Gohle, Udem, Herrmann,
  Rauschenberger, Holzwarth, Schuessler, Krausz, and
  H\"ansch}}]{gohle_frequency_2005}
\bibinfo{author}{\bibfnamefont{C.}~\bibnamefont{Gohle}},
  \bibinfo{author}{\bibfnamefont{T.}~\bibnamefont{Udem}},
  \bibinfo{author}{\bibfnamefont{M.}~\bibnamefont{Herrmann}},
  \bibinfo{author}{\bibfnamefont{J.}~\bibnamefont{Rauschenberger}},
  \bibinfo{author}{\bibfnamefont{R.}~\bibnamefont{Holzwarth}},
  \bibinfo{author}{\bibfnamefont{H.~A.} \bibnamefont{Schuessler}},
  \bibinfo{author}{\bibfnamefont{F.}~\bibnamefont{Krausz}}, \bibnamefont{and}
  \bibinfo{author}{\bibfnamefont{T.~W.} \bibnamefont{H\"ansch}},
  \bibinfo{journal}{Nature (London)} \textbf{\bibinfo{volume}{436}},
  \bibinfo{pages}{234} (\bibinfo{year}{2005}).

\bibitem[{\citenamefont{Jones et~al.}(2005)\citenamefont{Jones, Moll, Thorpe,
  and Ye}}]{jones_phase-coherent_2005}
\bibinfo{author}{\bibfnamefont{R.~J.} \bibnamefont{Jones}},
  \bibinfo{author}{\bibfnamefont{K.~D.} \bibnamefont{Moll}},
  \bibinfo{author}{\bibfnamefont{M.~J.} \bibnamefont{Thorpe}},
  \bibnamefont{and} \bibinfo{author}{\bibfnamefont{J.}~\bibnamefont{Ye}},
  \bibinfo{journal}{Phys. Rev. Lett.} \textbf{\bibinfo{volume}{94}},
  \bibinfo{pages}{193201} (\bibinfo{year}{2005}).

\bibitem[{\citenamefont{Benko et~al.}(2014)\citenamefont{Benko, Allison,
  Cing\"oz, Hua, Labaye, Yost, and Ye}}]{benko_extreme_2014}
\bibinfo{author}{\bibfnamefont{C.}~\bibnamefont{Benko}},
  \bibinfo{author}{\bibfnamefont{T.~K.} \bibnamefont{Allison}},
  \bibinfo{author}{\bibfnamefont{A.}~\bibnamefont{Cing\"oz}},
  \bibinfo{author}{\bibfnamefont{L.}~\bibnamefont{Hua}},
  \bibinfo{author}{\bibfnamefont{F.}~\bibnamefont{Labaye}},
  \bibinfo{author}{\bibfnamefont{D.~C.} \bibnamefont{Yost}}, \bibnamefont{and}
  \bibinfo{author}{\bibfnamefont{J.}~\bibnamefont{Ye}}, \bibinfo{journal}{Nat.
  Photonics} \textbf{\bibinfo{volume}{8}}, \bibinfo{pages}{530}
  (\bibinfo{year}{2014}).

\bibitem[{\citenamefont{Kandula et~al.}(2010)\citenamefont{Kandula, Gohle,
  Pinkert, Ubachs, and Eikema}}]{kandula_extreme_2010}
\bibinfo{author}{\bibfnamefont{D.~Z.} \bibnamefont{Kandula}},
  \bibinfo{author}{\bibfnamefont{C.}~\bibnamefont{Gohle}},
  \bibinfo{author}{\bibfnamefont{T.~J.} \bibnamefont{Pinkert}},
  \bibinfo{author}{\bibfnamefont{W.}~\bibnamefont{Ubachs}}, \bibnamefont{and}
  \bibinfo{author}{\bibfnamefont{K.~S.~E.} \bibnamefont{Eikema}},
  \bibinfo{journal}{Phys. Rev. Lett.} \textbf{\bibinfo{volume}{105}},
  \bibinfo{pages}{063001} (\bibinfo{year}{2010}).

\bibitem[{\citenamefont{Cing\"oz et~al.}(2012)\citenamefont{Cing\"oz, Yost,
  Allison, Ruehl, Fermann, Hartl, and Ye}}]{cingoz_direct_2012}
\bibinfo{author}{\bibfnamefont{A.}~\bibnamefont{Cing\"oz}},
  \bibinfo{author}{\bibfnamefont{D.~C.} \bibnamefont{Yost}},
  \bibinfo{author}{\bibfnamefont{T.~K.} \bibnamefont{Allison}},
  \bibinfo{author}{\bibfnamefont{A.}~\bibnamefont{Ruehl}},
  \bibinfo{author}{\bibfnamefont{M.~E.} \bibnamefont{Fermann}},
  \bibinfo{author}{\bibfnamefont{I.}~\bibnamefont{Hartl}}, \bibnamefont{and}
  \bibinfo{author}{\bibfnamefont{J.}~\bibnamefont{Ye}},
  \bibinfo{journal}{Nature (London)} \textbf{\bibinfo{volume}{482}},
  \bibinfo{pages}{68} (\bibinfo{year}{2012}).

\bibitem[{\citenamefont{Morgenweg et~al.}(2014)\citenamefont{Morgenweg, Barmes,
  and Eikema}}]{morgenweg_ramsey-comb_2014}
\bibinfo{author}{\bibfnamefont{J.}~\bibnamefont{Morgenweg}},
  \bibinfo{author}{\bibfnamefont{I.}~\bibnamefont{Barmes}}, \bibnamefont{and}
  \bibinfo{author}{\bibfnamefont{K.~S.~E.} \bibnamefont{Eikema}},
  \bibinfo{journal}{Nat. Phys.} \textbf{\bibinfo{volume}{10}},
  \bibinfo{pages}{30} (\bibinfo{year}{2014}).

\bibitem[{\citenamefont{Morgenweg and
  Eikema}(2014)}]{morgenweg_ramsey-comb_theory_2014}
\bibinfo{author}{\bibfnamefont{J.}~\bibnamefont{Morgenweg}} \bibnamefont{and}
  \bibinfo{author}{\bibfnamefont{K.~S.~E.} \bibnamefont{Eikema}},
  \bibinfo{journal}{Phys. Rev. A} \textbf{\bibinfo{volume}{89}},
  \bibinfo{pages}{052510} (\bibinfo{year}{2014}).

\bibitem[{\citenamefont{Ramsey}(1950)}]{ramsey_molecular_1950}
\bibinfo{author}{\bibfnamefont{N.~F.} \bibnamefont{Ramsey}},
  \bibinfo{journal}{Phys. Rev.} \textbf{\bibinfo{volume}{78}},
  \bibinfo{pages}{695} (\bibinfo{year}{1950}).

\bibitem[{\citenamefont{Altmann et~al.}(2016)\citenamefont{Altmann, Galtier,
  Dreissen, and Eikema}}]{altmann_high-precision_2016}
\bibinfo{author}{\bibfnamefont{R.~K.} \bibnamefont{Altmann}},
  \bibinfo{author}{\bibfnamefont{S.}~\bibnamefont{Galtier}},
  \bibinfo{author}{\bibfnamefont{L.~S.} \bibnamefont{Dreissen}},
  \bibnamefont{and} \bibinfo{author}{\bibfnamefont{K.~S.~E.}
  \bibnamefont{Eikema}}, \bibinfo{journal}{Phys. Rev. Lett.}
  \textbf{\bibinfo{volume}{117}}, \bibinfo{pages}{173201}
  (\bibinfo{year}{2016}).

\bibitem[{\citenamefont{Altmann et~al.}(2018)\citenamefont{Altmann, Dreissen,
  Salumbides, Ubachs, and Eikema}}]{altmann_deep-ultraviolet_2018}
\bibinfo{author}{\bibfnamefont{R.~K.} \bibnamefont{Altmann}},
  \bibinfo{author}{\bibfnamefont{L.~S.} \bibnamefont{Dreissen}},
  \bibinfo{author}{\bibfnamefont{E.~J.} \bibnamefont{Salumbides}},
  \bibinfo{author}{\bibfnamefont{W.}~\bibnamefont{Ubachs}}, \bibnamefont{and}
  \bibinfo{author}{\bibfnamefont{K.~S.~E.} \bibnamefont{Eikema}},
  \bibinfo{journal}{Phys. Rev. Lett.} \textbf{\bibinfo{volume}{120}},
  \bibinfo{pages}{043204} (\bibinfo{year}{2018}).

\bibitem[{\citenamefont{Corsi et~al.}(2006)\citenamefont{Corsi, Pirri, Sali,
  Tortora, and Bellini}}]{corsi_direct_2006}
\bibinfo{author}{\bibfnamefont{C.}~\bibnamefont{Corsi}},
  \bibinfo{author}{\bibfnamefont{A.}~\bibnamefont{Pirri}},
  \bibinfo{author}{\bibfnamefont{E.}~\bibnamefont{Sali}},
  \bibinfo{author}{\bibfnamefont{A.}~\bibnamefont{Tortora}}, \bibnamefont{and}
  \bibinfo{author}{\bibfnamefont{M.}~\bibnamefont{Bellini}},
  \bibinfo{journal}{Phys. Rev. Lett.} \textbf{\bibinfo{volume}{97}},
  \bibinfo{pages}{023901} (\bibinfo{year}{2006}).

\bibitem[{\citenamefont{Rundquist et~al.}(1998)\citenamefont{Rundquist,
  Durfee~III, Chang, Herne, Backus, Murnane, and
  Kapteyn}}]{rundquist_phase-matched_1998}
\bibinfo{author}{\bibfnamefont{A.}~\bibnamefont{Rundquist}},
  \bibinfo{author}{\bibfnamefont{C.~G.} \bibnamefont{Durfee~III}},
  \bibinfo{author}{\bibfnamefont{Z.}~\bibnamefont{Chang}},
  \bibinfo{author}{\bibfnamefont{C.}~\bibnamefont{Herne}},
  \bibinfo{author}{\bibfnamefont{S.}~\bibnamefont{Backus}},
  \bibinfo{author}{\bibfnamefont{M.~M.} \bibnamefont{Murnane}},
  \bibnamefont{and} \bibinfo{author}{\bibfnamefont{H.~C.}
  \bibnamefont{Kapteyn}}, \bibinfo{journal}{Science}
  \textbf{\bibinfo{volume}{280}}, \bibinfo{pages}{1412} (\bibinfo{year}{1998}).

\bibitem[{\citenamefont{Popmintchev et~al.}(2009)\citenamefont{Popmintchev,
  Chen, Bahabad, Gerrity, Sidorenko, Cohen, Christov, Murnane, and
  Kapteyn}}]{popmintchev_phase_2009}
\bibinfo{author}{\bibfnamefont{T.}~\bibnamefont{Popmintchev}},
  \bibinfo{author}{\bibfnamefont{M.-C.} \bibnamefont{Chen}},
  \bibinfo{author}{\bibfnamefont{A.}~\bibnamefont{Bahabad}},
  \bibinfo{author}{\bibfnamefont{M.}~\bibnamefont{Gerrity}},
  \bibinfo{author}{\bibfnamefont{P.}~\bibnamefont{Sidorenko}},
  \bibinfo{author}{\bibfnamefont{O.}~\bibnamefont{Cohen}},
  \bibinfo{author}{\bibfnamefont{I.~P.} \bibnamefont{Christov}},
  \bibinfo{author}{\bibfnamefont{M.~M.} \bibnamefont{Murnane}},
  \bibnamefont{and} \bibinfo{author}{\bibfnamefont{H.~C.}
  \bibnamefont{Kapteyn}}, \bibinfo{journal}{Proc. Natl. Acad. Sci.}
  \textbf{\bibinfo{volume}{106}}, \bibinfo{pages}{10516}
  (\bibinfo{year}{2009}).

\bibitem[{\citenamefont{Sali\`eres et~al.}(1999)\citenamefont{Sali\`eres,
  Le~D\'eroff, Auguste, Monot, d'Oliveira, Campo, Hergott, Merdji, and
  Carr\'e}}]{salieres_frequency-domain_1999}
\bibinfo{author}{\bibfnamefont{P.}~\bibnamefont{Sali\`eres}},
  \bibinfo{author}{\bibfnamefont{L.}~\bibnamefont{Le~D\'eroff}},
  \bibinfo{author}{\bibfnamefont{T.}~\bibnamefont{Auguste}},
  \bibinfo{author}{\bibfnamefont{P.}~\bibnamefont{Monot}},
  \bibinfo{author}{\bibfnamefont{P.}~\bibnamefont{d'Oliveira}},
  \bibinfo{author}{\bibfnamefont{D.}~\bibnamefont{Campo}},
  \bibinfo{author}{\bibfnamefont{J.-F.} \bibnamefont{Hergott}},
  \bibinfo{author}{\bibfnamefont{H.}~\bibnamefont{Merdji}}, \bibnamefont{and}
  \bibinfo{author}{\bibfnamefont{B.}~\bibnamefont{Carr\'e}},
  \bibinfo{journal}{Phys. Rev. Lett.} \textbf{\bibinfo{volume}{83}},
  \bibinfo{pages}{5483} (\bibinfo{year}{1999}).

\bibitem[{\citenamefont{Chen et~al.}(2007)\citenamefont{Chen, Varma, Alexeev,
  and Milchberg}}]{chen_measurement_2007}
\bibinfo{author}{\bibfnamefont{Y.-H.} \bibnamefont{Chen}},
  \bibinfo{author}{\bibfnamefont{S.}~\bibnamefont{Varma}},
  \bibinfo{author}{\bibfnamefont{I.}~\bibnamefont{Alexeev}}, \bibnamefont{and}
  \bibinfo{author}{\bibfnamefont{H.}~\bibnamefont{Milchberg}},
  \bibinfo{journal}{Opt. Express} \textbf{\bibinfo{volume}{15}},
  \bibinfo{pages}{7458} (\bibinfo{year}{2007}).

\bibitem[{\citenamefont{Morgenweg and Eikema}(2012)}]{morgenweg_tailored_2012}
\bibinfo{author}{\bibfnamefont{J.}~\bibnamefont{Morgenweg}} \bibnamefont{and}
  \bibinfo{author}{\bibfnamefont{K.~S.~E.} \bibnamefont{Eikema}},
  \bibinfo{journal}{Opt. Lett.} \textbf{\bibinfo{volume}{37}},
  \bibinfo{pages}{208} (\bibinfo{year}{2012}).

\bibitem[{\citenamefont{Morgenweg and
  Eikema}(2013)}]{morgenweg_multi-delay_2013}
\bibinfo{author}{\bibfnamefont{J.}~\bibnamefont{Morgenweg}} \bibnamefont{and}
  \bibinfo{author}{\bibfnamefont{K.~S.~E.} \bibnamefont{Eikema}},
  \bibinfo{journal}{Opt. Express} \textbf{\bibinfo{volume}{21}},
  \bibinfo{pages}{5275} (\bibinfo{year}{2013}).

\bibitem[{\citenamefont{W\"ormer et~al.}(1989)\citenamefont{W\"ormer,
  Guzielski, Stapelfeldt, and M\"oller}}]{wormer_fluorescence_1989}
\bibinfo{author}{\bibfnamefont{J.}~\bibnamefont{W\"ormer}},
  \bibinfo{author}{\bibfnamefont{V.}~\bibnamefont{Guzielski}},
  \bibinfo{author}{\bibfnamefont{J.}~\bibnamefont{Stapelfeldt}},
  \bibnamefont{and} \bibinfo{author}{\bibfnamefont{T.}~\bibnamefont{M\"oller}},
  \bibinfo{journal}{Chem. Phys. Lett.} \textbf{\bibinfo{volume}{159}},
  \bibinfo{pages}{321} (\bibinfo{year}{1989}).

\bibitem[{\citenamefont{Ozawa and Kobayashi}(2013)}]{ozawa_vuv_2013}
\bibinfo{author}{\bibfnamefont{A.}~\bibnamefont{Ozawa}} \bibnamefont{and}
  \bibinfo{author}{\bibfnamefont{Y.}~\bibnamefont{Kobayashi}},
  \bibinfo{journal}{Phys. Rev. A} \textbf{\bibinfo{volume}{87}},
  \bibinfo{pages}{022507} (\bibinfo{year}{2013}).

\bibitem[{\citenamefont{Chan et~al.}(1992)\citenamefont{Chan, Cooper, Guo,
  Burton, and Brion}}]{chan_absolute_1992}
\bibinfo{author}{\bibfnamefont{W.~F.} \bibnamefont{Chan}},
  \bibinfo{author}{\bibfnamefont{G.}~\bibnamefont{Cooper}},
  \bibinfo{author}{\bibfnamefont{X.}~\bibnamefont{Guo}},
  \bibinfo{author}{\bibfnamefont{G.~R.} \bibnamefont{Burton}},
  \bibnamefont{and} \bibinfo{author}{\bibfnamefont{C.~E.} \bibnamefont{Brion}},
  \bibinfo{journal}{Phys. Rev. A} \textbf{\bibinfo{volume}{46}},
  \bibinfo{pages}{149} (\bibinfo{year}{1992}).

\bibitem[{\citenamefont{Armstrong et~al.}(1997)\citenamefont{Armstrong, Alford,
  Raymond, Smith, and Bowers}}]{armstrong_parametric_1997}
\bibinfo{author}{\bibfnamefont{D.~J.} \bibnamefont{Armstrong}},
  \bibinfo{author}{\bibfnamefont{W.~J.} \bibnamefont{Alford}},
  \bibinfo{author}{\bibfnamefont{T.~D.} \bibnamefont{Raymond}},
  \bibinfo{author}{\bibfnamefont{A.~V.} \bibnamefont{Smith}}, \bibnamefont{and}
  \bibinfo{author}{\bibfnamefont{M.~S.} \bibnamefont{Bowers}},
  \bibinfo{journal}{J. Opt. Soc. Am. B} \textbf{\bibinfo{volume}{14}},
  \bibinfo{pages}{460} (\bibinfo{year}{1997}).

\bibitem[{\citenamefont{Yoshino and Freeman}(1985)}]{yoshino_absorption_1985}
\bibinfo{author}{\bibfnamefont{K.}~\bibnamefont{Yoshino}} \bibnamefont{and}
  \bibinfo{author}{\bibfnamefont{D.~E.} \bibnamefont{Freeman}},
  \bibinfo{journal}{J. Opt. Soc. Am. B} \textbf{\bibinfo{volume}{2}},
  \bibinfo{pages}{1268} (\bibinfo{year}{1985}).

\bibitem[{\citenamefont{Cadoret et~al.}(2009)\citenamefont{Cadoret,
  De~Mirandes, Clad{\'e}, Nez, Julien, Biraben, and
  {Guellati-Kh{\'e}lifa}}}]{cadoret_atom_2009}
\bibinfo{author}{\bibfnamefont{M.}~\bibnamefont{Cadoret}},
  \bibinfo{author}{\bibfnamefont{E.}~\bibnamefont{De~Mirandes}},
  \bibinfo{author}{\bibfnamefont{P.}~\bibnamefont{Clad{\'e}}},
  \bibinfo{author}{\bibfnamefont{F.}~\bibnamefont{Nez}},
  \bibinfo{author}{\bibfnamefont{L.}~\bibnamefont{Julien}},
  \bibinfo{author}{\bibfnamefont{F.}~\bibnamefont{Biraben}}, \bibnamefont{and}
  \bibinfo{author}{\bibfnamefont{S.}~\bibnamefont{{Guellati-Kh{\'e}lifa}}},
  \bibinfo{journal}{The European Physical Journal Special Topics}
  \textbf{\bibinfo{volume}{172}}, \bibinfo{pages}{121} (\bibinfo{year}{2009}).

\end{thebibliography}
\end{document}